# IMPROVED PHYSICAL PROPERTIES DATABASE OF TRANS-NEPTUNIAN DWARF PLANETS

Yury I. Rogozin

Veda LLC, Moscow, Russia; e-mail: yrogozin@gmail.com

**ABSTRACT**

Because of the objectively existing causes, such as an irregular shape of a celestial body, the lack of the satellites, the low values of albedo at moderate sizes, and a large remoteness from the Sun, creating a more exact the physical properties database of trans-Neptunian dwarf planets in comparison with existing that continues to remain a problem. We offer a new calculation procedure of the physical properties of these objects involving earlier unknown the relationships harmonizing these properties. It allows us to conjecture of they are a group of uniform celestial bodies by their origin beyond the Kuiper belt. The calculated physical properties of these dwarf planets are in a good agreement with the estimates received from observational data and can form the basis for support the validity of this improved physical properties database.

## 1. INTRODUCTION

In the early 21-st century planetary science has been discovered beyond Neptune's orbit several the largest trans-Neptunian objects (TNOs). As is known, after excluding Pluto from among classical planets the General assembly of International Astronomical Union in 2006 has entered a new category of celestial bodies named "dwarf planets". In this category, besides Pluto, such three recently discovered the largest TNOs as 2003 UB313, 2003 EL61, and 2005FY9 then received new names accordingly as Eris, Haumea, and Makemake were included. One of the distinctive features of these trans-Neptunian dwarf planets (TNDPs) is their unusually high value of inclination (more than $15^0$). Besides an appreciable dichotomy exist in the observable values of albedo and spectral characteristics of the largest TNOs on the on hand, and the rest, smaller objects of the Kuiper belt on the other. Apparently, proceeding from this, the assumption recently was put forward that "they are members of a unique physical class within the Kuiper belt population" (Stansberry et al 2008). In support of this assumption could testify as well an availability of a harmony of the orbital parameters and physical properties of these TNDPs. As it is well known, as against the classical planets of the Solar System the orbital parameters of these TNDPs have been not systematized to date and had not a unified rule for orbital distances. As to the available now the physical properties database of these TNDPS, it does not allow with a confidence to establish an availability of such harmony as suffers for noticeable uncertainty of these properties at one degree or another practically with all known TNDPs. The properties even of the long-explored Pluto have some ambiguity, namely mass $(1.305 \pm 0.007) \times 10^{21}$ kg and $(1.309 \pm 0.018) \times 10^{21}$ kg, radius $1,153 \pm 10$ km and $1,151 \pm 6$ km, and density $2.03 \pm 0.06$ g cm$^{-3}$ and $2.05 \pm 0.04$ g cm$^{-3}$, respectively, in the former case (Buie et al 2006) and the second (http://ssd.jpl.nasa.gov/?planet_phys_par). The estimates of Eris' radius fall in the broad range between $1,163 \pm 6$ km (Sicardi et al 2011) and $1,500 \pm 200$ km (Bertoldi et al 2006). At Eris' mass value $(1.66 \pm 0.02) \times 10^{22}$ kg (Brown & Schaller 2007) the respective scatter in its density's estimates is from 2.52 g cm$^{-3}$ to 1.2 g cm$^{-3}$. Other widely known estimate of radius and density of Eris is $1,200 \pm 50$ km and (Brown et al 2006) and $2.3 \pm 0.3$ g cm$^{-3}$ (Brown & Schaller 2007).

The density of the third dwarf planet Haumea because of complexity its odd shape is estimated only approximately as 2.6 ÷ 3.3 g cm$^{-3}$. As to the density of one more dwarf planet Makemake, it is considered approximately equal 2 g cm$^{-3}$ as exact value of its mass remains as yet unknown. Besides this this database, probably, could be filled up at the expense of available candidates in TNDPs but their data, e.g. the largest TNO 2002 TC302 suffer of an absence the real data about its mass and density. Thus, it is possible to conclude that at least now alone photometric or radiometric data of these dwarf planets it appears not sufficiently to give a reasonably accurate and full idea about their physical properties (size, mass, density, and surface gravity).

Therefore, a purpose of this letter is refinement of the existing physical properties database of TNDPs using previously unknown harmonic relationships between the physical properties of these dwarf planets.

## 2. CALCULATION PROCEDURE

At present both the orbital parameters and the physical properties data of each TNDP exist by itself. Yet, as we shall see subsequently they are related together through mathematical relationships. For this purpose we shall to trace a behavior for any property at least for three TNDPs. As yet the physical properties of as little as two TNDPs, Pluto and Eris, with some degree of certainty are known in full measure. The cornerstone of present calculation procedure is the need for use of the sample physical data of Charon as third component of such dependence. Whilst at present Charon isn't a self-consistent TNO, but it could likely to be one in the past. This follows from a rule for the orbital distances of TNDPs that we find by the trial-and-error method:

$$R_n = a\,[1 + (n-1)\,b]\,R_N, \qquad (1)$$

where the coefficients $a = \Phi^2/2$ and $b = 1/4\pi$ with $\Phi = 1.6180$ and $\pi = 3.1416$ – fundamental mathematical constants, n – serial number of TNO, and $R_n$ and $R_N$ are semi-major axes of orbits accordingly for n-th largest TNO and Neptune. This formula gives $R_n$ = 39.36 AU; 42.49 AU; 45.63 AU, and 67.55 AU respectively for Pluto (n=1), Haumea (n=2), Makemake (n=3), and Eris (n=10) at observational $R_n$ respectively 39.48 AU; 43.13 AU; 45.79 AU, and 67.67 AU. Here, it is significant that according to this rule there could be the orbit with n = 0 and $R_n$ = 36..22 AU which most likely has been belonged to Charon that quite possible was captured by Pluto considering its perihelion is less than 30 AU. Really, difference between the orbital speed of Charon on its former possible orbit and present orbital speed of Pluto/Charon system is 0.209 km s$^{-1}$ that is close to present orbital speed of Charon that equal 0.224 km s$^{-1}$. The foregoing orbital data and the mean value of Charon's mass $M_C \approx 1.60 \times 10^{21}$ kg accepted as 0.122 of Pluto's mass $M_p$ (Olkin et al 2003) made possible to establish that the dependence of their surface gravity $g_n$ of the angular momentum is of the curve of logarithmic function $g_n = a\,\ln(M_n^{*} v_n R_n) + b$, where $M_n^{*}$, $v_n$, and $R_n$ are respectively mass, orbital speed, and semi-major axis of the n-th TNDP. We have chosen from these three objects Pluto and Charon to derive a necessary dependence in view of smaller validity of Eris for this purpose because of a foregoing wide scatter of its size estimates. For the sake of convenience of calculations at a search for the specific form of this function instead of the valid value of mass of these objects $M_n$ in kg its normalized value $M_n^{*}$ (= $M_n/10^{21}$) here is used, and orbital speed $v_n$ and semi-major axis $R_n$ are used in dimensions respectively km s$^{-2}$ and AU . With data of 13.05 x 10$^{21}$ kg, 4.666 km s$^{-1}$, and 39.48 AU for Pluto and 1.60 x $10^{21}$ kg, 4.95 km s$^{-1}$, and 36.22 AU for Charon their angular momentums in these conventional units are respectively 2,404 for Pluto and 287 for Charon. In doing so with mean radius of Pluto 1,153 km (Buie et al 2006) and that of Charon 593 km (Olkin et al 2003) their surface gravities makes up 0.655 m s$^{-2}$ and 0.304 m s$^{-2}$ respectively. From this system of two equations such values follows: $a = 0.1651$ and $b = -0.6303$, i.e.

$$g_n = a \ln (M_n^* v_n R_n) - b = 0.1651 \ln (M_n^* v_n R_n) - 0.6303 \qquad (2)$$

Thereupon, substituting the value of surface gravity $g_n$ of n-th TNDP obtained from Eq. (2) in known expression $\qquad g\, r^2 = G\, M, \qquad (3)$
where $G$ is Newton's constant, we can derive mean radius $r$ of n-th TNDP with known mass.

## 3. RESULTS

First we shall use the foregoing procedure to debatable size of Eris. From Eq.(2) it is evident that its $g = 0.7332$ m s$^{-2}$. As a result Eris' radius makes up 1,229 km or its diameter is 2,458 km. Very nearly the same value 2,455 km accordingly to this procedure has been obtained previously (Rogozin 2009). This small difference in diameter of Eris is due to slightly distinct used physical properties of Pluto and Charon. Although result about $D = 2,458$ km is compatible with recent size/albedo estimate for Eris/Dysnomia system $D = 2,454 \pm 117$ km (Santos-Sanz et al 2012) and appears to be more plausible than the foregoing estimates, we believe that it can be just refined. Numerical values for coefficients "$a$" and "$b$" in Eq. (2) may slightly vary with particular used physical properties data of Pluto and Charon. To obtain more general form of such relationship we may anticipate that just like many physical laws these two coefficients can be expressed through fundamental mathematical constants. Evidently in this instance $a = 0.1651 \approx \Phi/\pi^2 = 0.16394$ and $b = -0.6303 \approx -1/\Phi = -0.61803$. With these substitutions, Eq. (2) becomes

$$g_n = (\Phi/\pi^2) \ln (M_n^* v_n R_n) - 1/\Phi \qquad (2a)$$

With use the Eris' mass $(16.6 \pm 0.2)$ x $10^{21}$ kg and its $v_n = 3.436$ km s$^{-1}$ in view of $g_n = 0.73584 \pm 0.0200$ m s$^{-2}$ from Eq.(2a) we have its mean radius $1,227 \pm 6$ km or its mean $D = 2,454 \pm 12$ km that surprisingly exactly agree with the cited recent estimate (Santos-Sanz et al 2012). In doing so Eris' density make up $2.145 +0.006/-0.004$ g cm$^{-3}$, that is compatible within foregoing error bar $2.3 \pm 0.3$ g cm$^{-3}$. On the assumption that Eris has non-spherical shape its diameter deviations from the mean can increase about $\pm 100$ km.
In a similar way in the example of TNDP Haumea using its mass value $(4.006 \pm 0.040)$ x $10^{21}$ kg (Ragozzine & Brown 2009) in view of $g_n = 0.4744 \pm 0.0016$ m s$^{-2}$ we have its mean diameter $1,501 \pm 5$ km and its density $2.26 \pm 0.04$ g cm$^{-3}$.
Yet, this approach isn't adequate for TNDP Makemake as its mass as yet is unknown. Because of this we used recent measurements of its diameter $1,420 \pm 60$ km (Lim et al 2012). In doing so its mass value obtainable from Eqs. (2a) and (3) make up $(3.355 + 0.545/-0.355)$ x $10^{21}$ kg. Respectively its density is about $2.24 \div 2.30$ g cm$^{-3}$. To reach a good agreement between $g$ – values in Eqs. (2a) and (3) as the mean diameter value of Makemake was adopted 1,412 km and at this condition its mass is found to be $3.367$ x $10^{21}$ kg with its density equal 2.28 g cm$^{-3}$ and $g$ equal 0.450 m s$^{-2}$.
One further the largest TNO 2002 TC302 fit Eq. (1) at number 6. Indeed, from this equation its $R_n$ equals 55.02 AU versus observable value 55.24 AU. By analogy with TNDPs Eqs. (2a) and

(3) is useful as applied also to this TNO. In doing so we used *Spitzer* data about the mean diameter of 2002 TC302 1,150 km. There from these equations follows its mass to be equal 1.81 x $10^{21}$ kg. Respectively its density is 2.27 g cm$^{-3}$ and its surface gravity is 0.364 m s$^{-2}$. Finally, we shall return to Pluto and Charon to refine their physical properties using Eq. (2a). The closest possible fit between *g*-values found from Eqs. (2a) and (3) for Pluto in the case of its mass value 13.05 x $10^{21}$ kg is attained at its radius 1,150.5 km resulting its density 2.046 g cm$^{-3}$, which lies within foregoing error bar 2.05 ± 0.04 g cm$^{-3}$. In similar manner this approach for Charon with its mass 1.60 x $10^{21}$ kg gives radius 587 km and its density 1.89 g cm$^{-3}$. Here, we would like point out one remarkable fit, namely, a sample equality of ratios for known mean values of "occultation" radii of Pluto 1,180 km (Millis et al 1993) and Charon 603 km (Sicardi et al 2006) to their mean found radii respectively 1,150.5 km and 587 km, that is one further manifestation of harmony of these related celestial bodies. An attempt to apply this ratio ≈ 1.026 to Eris as a twin of Pluto with the similar physical properties and composition using its recent "occultation" diameter 2,326 km (Sicardi et al 2011) gives its mean diameter 2,267 km that is even less than the lower photometric estimate 2,300 km (Brown et al 2006) and show up to be quite unbelievable.

To gain a better perception of found improved physical properties for known TNDPs, Charon, and TNO 2002 TC302 their summary gives in Table 1. Here, for simplicity the mean values of the physical properties for these objects without considering their uncertainties are listed.

Table 1: Improved physical properties for TNDPs, Charon, and TNO 2002 TC302

| Parameters | Eris | Pluto | Haumea | Makemake | Charon | TC302 |
|---|---|---|---|---|---|---|
| Mean diameter, km | 2,454 | 2,301 | 1,501 | 1,412 | 1,174 | 1,150 |
| Density, g cm$^{-3}$ | 2.145 | 2.046 | 2.26 | 2.28 | 1.89 | 2.27 |
| Mass, kg (x$10^{21}$) | 16.6 | 13.05 | 4.006 | 3.367 | 1.60 | 1.81 |
| Surface gravity, m s$^{-2}$ | 0.736 | 0.658 | 0.474 | 0.450 | 0.310 | 0.364 |

## 4. CONCLUSION

The distinct features of our calculation procedure of improved physical properties for TNDPs is using the temporary Charon's orbital data as a former self-consistent large TNO that can be lost by Neptune among other its satellites now named trans-Neptunian dwarf planets, and previously unknown relationship for their surface gravity. The validity of this approach is supported by the exact agreement between the foregoing independent Eris' diameter estimates of 2,454 km and good fit of improved physical properties for other TNDPs to their observational data. Regarding other recent Eris' diameter estimates may be supposed that they are subject to some neglected factors. In particular, its mean diameter value 2,509 km (Rogozin 2012) has been slightly overestimated considering that a direct application of the empirical coefficient of proportionality C = π adopted for Pluto/Charon system in Eq. (2) of eccentricity in that work to other TNDPs appears isn't universally true as it fail to account for some unknown factors possibly related to their strongly different orbital parameters and the distinction Pluto/Charon system as binary from the remaining TNDPs.

Thus, as against an existing set of isolated physical properties data for TNDPs we have received the concordant system of such properties complemented by respective data for TNO 2002 TC302 as possible further TNDP. As it is follows from a good agreement between the quoted recent observational data about the Eris' size (Santos-Sanz et al 2012) and our calculation data the superiority of Eris' size over Pluto's consists more than 6%. Because of this a moot point "what is the biggest dwarf planet of the two (Pluto or Eris)" appears may be thought of closed.

*Note*: The shown here a harmony of the orbital distances and the physical properties of TNDPs (including Charon), the existence of their mean-motion resonances with Neptune, and also the close values of density indicating of their structural similarity, apparently, testify to their particular origin beyond the Kuiper belt as the possible former Neptune's satellites that could be lost at its entry in the Solar System as newcomer. Thus, on the basis such their prospective origin these TNDPs hardly can be considered as some specific category of planets. In our opinion, these discovered TNDPs and related with them TNO 2002 TC302 would be more properly referred to as plansats (derived from words "planet" and "satellite"). In connection with this possibly it would be well to replace a traditional way of assignment the names to such new objects of the Solar System unlike classical planets by the names in honor of world-wide scientists in field of astronomy and astrophysics, such as Copernicus, Kepler, Galilei, and Newton (e.g. Copra, Kepla, Galila, and Newta).